# Multiple magnetoplasmon polaritons of magneto-optical graphene in near-field radiative heat transfer

Ming-Jian He[1,2], Lei Qu[1,2], Ya-Tao Ren[1,2,*], Hong Qi[1,2,**], Mauro Antezza[3,4], and He-Ping Tan[1,2]

1 School of Energy Science and Engineering, Harbin Institute of Technology, Harbin 150001, P. R. China

2 Key Laboratory of Aerospace Thermophysics, Ministry of Industry and Information Technology, Harbin 150001, P. R. China

3 Laboratoire Charles Coulomb (L2C), UMR 5221 CNRS-Université de Montpellier, F-34095 Montpellier, France

4 Institut Universitaire de France, 1 rue Descartes, F-75231 Paris, France

*Corresponding author: Email: renyt@hit.edu.cn (YT. Ren); qihong@hit.edu.cn (H Qi)

**Abstract:** Graphene, as a two-dimensional magneto-optical material, supports magnetoplasmon polaritons (MPP) when exposed to an applied magnetic field. Recently, MPP of a single-layer graphene has shown an excellent capability in the modulation of near-field radiative heat transfer (NFRHT). In this study, we present a comprehensive theoretical analysis of NFRHT between two multilayered graphene structures, with a particular focus on the multiple MPP effect. We reveal the physical mechanism and evolution law of the multiple MPP, and we demonstrate that the multiple MPP allow one to mediate, enhance, and tune the NFRHT by appropriately engineering the properties of graphene, the number of graphene sheets, the intensity of magnetic fields, as well as the geometric structure of systems. We show that the multiple MPP have a quite significant distinction relative to the single MPP or multiple surface plasmon polaritons (SPPs) in terms of modulating and manipulating NFRHT. We demonstrate that this remarkable behavior is attributed to the coupling between the significant contributions of surface states at multiple surfaces and Shubnikov–de Haas-like oscillations in the spectrum, indicating a transformation of intraband and interband transitions. Notably, we find that the evolution from single MPP to multiple MPP is absolutely different from that from single SPPs to multiple SPPs. Finally, a giant thermal magnetoresistance effect and a negative-positive transition of the relative thermal magnetoresistance ratio are predicted in the multilayered system under consideration. Our study paves the way for a flexible control of NFRHT and it offers the possibility for the thermal photon-based communication technology and a magnetically controllable thermal switch.

**Keywords:** near-field radiative heat transfer, magneto-optical graphene, multiple magnetoplasmon polaritons,

thermal magnetoresistance

# I. INTRODUCTION

Miniaturization has emerged as a significant trend in scientific development, particularly in the realm of natural sciences. Gaining a comprehensive understanding and mastery of thermal behavior and mechanisms at the nanoscale has become an imperative for the in-depth advancement of heat transfer. It has been proved that the radiative heat flux can surpass Planck's blackbody limit by several orders of magnitude, when the separation distance between the two bodies is comparable to or smaller than the characteristic wavelength of the thermal radiation [1-4]. This unique physical phenomenon is commonly referred to as near-field radiative heat transfer (NFRHT) [5-7]. In recent decades, experimental validations [8-10] have confirmed the theory of NFRHT, leading to significant advancements in theoretical findings [11-15].

Flexible control of NFRHT has consistently captivated researchers due to its potential to revolutionize thermal management in functional devices, such as integrated circuits and near-field thermophotovoltaics [10, 16-18]. The control and modulation of NFRHT can be classified into two categories: passive methods, which rely on the properties of materials and the geometric parameters of the structure [19-23], and active control methods, which involve mechanical actions such as rotation and strain [24-26]. In contrast to mechanical actions that may cause irreversible damage to the structure, a promising and emerging active control method involves the application of external fields, such as electric or magnetic fields [27-31]. This method offers a flexible approach to modulating NFRHT [32]. In this respect, magneto-optical semiconductors, such as InSb or Si, have shown remarkable performance in modulating NFRHT, primarily due to their unique optical properties that are sensitive to magnetic fields [33-35]. Plasmonic structures made of these conventional magnetic materials have exhibited giant thermal magnetoresistance and nonreciprocal surface modes [36-40].

Graphene, exhibits intriguing electronic properties due to the massless Dirac dispersion of electrons. Due to its highly confined surface plasmon polaritons (SPPs), graphene has generated significant interest in the field of thermal photonics [41-44]. It has demonstrated exceptional modulation capabilities in NFRHT [45-48], thereby contributing to the advancement of novel thermally functional devices [49-51]. Interestingly, recent discoveries have revealed that when subjected to an external magnetic field, graphene's conductivity becomes a tensor. It exhibits finite values for both diagonal and off-diagonal components in the Hall regime. Hybridization occurs between cyclotron excitations and plasmons in magneto-optical graphene, giving rise to magnetoplasmon polaritons (MPP) [52]. In MPP, the original continuous Dirac energy spectrum of SPPs transforms into distinct

Landau levels, leading to the emergence of unique phenomena such as quantum Faraday effects and Kerr effects [52].

Building upon the quantum Hall regime of graphene, researchers have observed intriguing Shubnikov–de Haas-like oscillations in the spectral radiative heat flux between two suspended magneto-optical graphene sheets. These oscillations arise from the intraband and interband transitions among different Landau levels [53, 54]. Subsequently, the active manipulation of NFRHT using MPP was reported between two suspended graphene gratings through the combination of magnetic field and twisting action [55]. Currently, there is growing interest in the magnetic field dependence of NFRHT between two graphene-based hyperbolic metamaterials, such as those composed of alternating layers of graphene and polar dielectric materials [56, 57].

The strong interaction between graphene and polar materials in the aforementioned structures leads to the occurrence of hybridization modes between multiple MPP and surface phonon polaritons. These surface phonon polaritons are electromagnetic waves that propagate along the interfaces of polar dielectrics, displaying a significant local-field enhancement in the proximity of the interfaces. However, the presence of these hybridization modes complicates the comprehensive understanding of the physical mechanism and evolutionary behavior of pure multiple MPP in NFRHT. These phenomena exhibit an exceptionally intricate nature in the realm of physics due to the strong coupling characteristics of multiple modes and the complex nature of multilayered structures. Their complexity is determined by various parameters, including the properties of graphene (chemical potentials), the number of graphene sheets, the intensity of the magnetic field, and the geometric parameters of the system (such as the interlayer distance of graphene). Consequently, the NFRHT mediated by multiple MPP can be significantly influenced by all the above parameters.

In fact, multilayer structures are excellent candidates for enhancing NFRHT due to the significant contributions of surface states at multiple surfaces [58]. However, despite the aforementioned reasons, the mechanism and evolutionary behavior of multiple MPP induced by multilayered magneto-optical graphene sheets, as well as their manipulation in NFRHT, remain unclear. Understanding these aspects is of great significance in the field of physics and holds the potential to open up new avenues for flexible control of NFRHT. This study aims to systematically investigate the NFRHT between multilayered graphene structures in the presence of a magnetic field, with a specific focus on multiple MPP. We comprehensively explore and investigate the mechanism, evolution, and manipulation of multiple MPP in NFRHT.

The paper is structured as follows: In Section II, the physical system is introduced by giving the magneto-optical conductivity of graphene and the formulas of NFRHT for multilayered anisotropic 2D layers. Section III focuses on the investigation of the modulation of multiple MPP in NFRHT and explores and demonstrates the physical mechanism behind their manipulation. This section also systematically considers all the aforementioned factors. Lastly, Section IV presents the conclusive remarks.

## II. PHYSICAL SYSTEM

By applying a perpendicular magnetic field to graphene, the response of graphene electrons to the external magnetic field exhibits a characteristic optical quantum Hall effect. The conductivity of graphene transforms into a tensor with nonzero values for both diagonal and off-diagonal components. This is described by [59, 60]

$$\begin{pmatrix} \sigma_{xx} & \sigma_{xy} \\ \sigma_{yx} & \sigma_{yy} \end{pmatrix} = \begin{pmatrix} \sigma_L & \sigma_H \\ -\sigma_H & \sigma_L \end{pmatrix} \tag{1}$$

where $\sigma_L$ and $\sigma_H$ represent the longitudinal and Hall conductivities, respectively. Within the Dirac-cone approximation, the magneto-optical conductivity of graphene takes on a simple form in the random phase approximation [52]

$$\sigma_{L(H)}(\omega, B) = g_s g_v \times \frac{e^2}{4h} \sum_{n \neq m = -\infty}^{+\infty} \frac{\Xi_{L(H)}^{nm}}{i\Delta_{nm}} \frac{n_F(E_n) - n_F(E_m)}{\hbar\omega + \Delta_{nm} + i\Gamma_{nm}(\omega)} \tag{2}$$

where $g_{s(v)}=2$ is the spin (valley) degeneracy factor of graphene, $e$ is the charge of an electron, and $h$ is Planck's constant. $n_F(E_n) = 1/\left[1 + e^{(E_n - \mu)/k_B T}\right]$ stands for the Fermi distribution function, $k_B$ is the Boltzmann constant, and $\mu$ and $T$ are the chemical potential and temperature of graphene, respectively. $\Gamma_{nm}(\omega)$ is the LL broadening, taken as 6.8 meV in the present work.

With the application of an external magnetic field, electrons acquire significant cyclotron energies due to the Lorentz force. The continuous Dirac energy spectrum undergoes a transformation into degenerate Landau levels (LLs), where the energy of the $n$-th LL is given by [52]

$$E_n = \text{sign}(n)(\hbar v_F / l_B)\sqrt{2|n|} \tag{3}$$

where $v_F \approx 10^6$ m/s is the Fermi velocity of the carriers in graphene, $l_B = \sqrt{\hbar/eB}$ is the magnetic length, and $B$ is the intensity of the magnetic field. $\Delta_{nm} = E_n - E_m$ represents the energy transition between Landau levels ($m$, $n$= 0, ±1, ±2, . . .), where the LL indices indicate the respective Landau levels. The matrix elements in Eq. (2) are given by [52]

$$\Xi_L^{nm} = \frac{\hbar^2 v_F^2}{l_B^2}\left(1 + \delta_{m,0} + \delta_{n,0}\right)\delta_{|n|-|m|,\pm 1}$$
$$\Xi_H^{nm} = i\Xi_L^{nm}(\delta_{|m|,|n|-1} - \delta_{|m|-1,|n|}) \quad (4)$$

where $\delta$ is the Kronecker symbol. The band structure of graphene is composed of two Dirac cones. The number of last occupied LLs is denoted as $N_F$=int[$(\mu/E_1)^2$], which represents the number of occupied electron-degenerate Landau levels. Electronic transitions in magneto-optical graphene can be categorized into two types: (i) intraband transitions occurring between adjacent Landau levels (from $N_F$ to $N_F$+1) within the positive cone; and (ii) interband transitions that connect Landau levels from the negative cone with Landau levels in the positive cone.

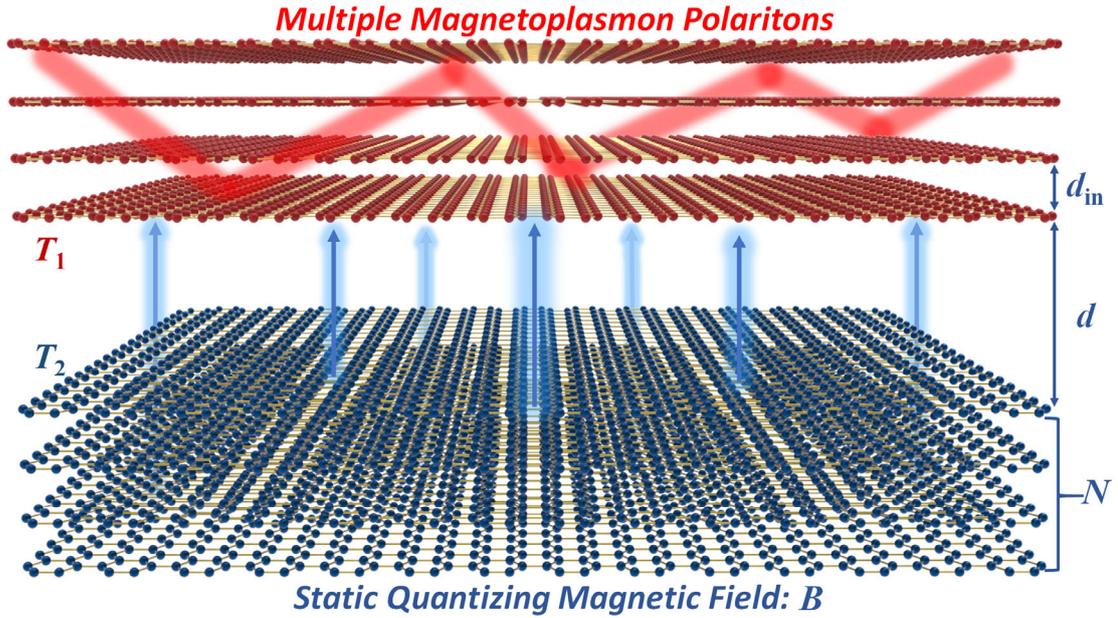

Fig. 1. Schematic of near-field radiative heat transfer between two bodies (kept at temperatures $T_1$ and $T_2$) composed of multilayered graphene sheets (the number of the sheets is identical for the two bodies, indicated as $N$). The system is assumed to be localized in vacuum, and the separation distances between the two bodies and the adjacent graphene sheets are denoted as $d$ and $d_{in}$, respectively. $d$ and $d_{in}$ are both selected as 10 nm, if no otherwise specified. A static quantizing magnetic field $B$ is applied perpendicularly to the system.

In this study, we examine the near-field radiative heat transfer between two bodies consisting of multiple layers of graphene sheets, and the number of the sheets is identical for the two bodies, labeled as *N* in Fig. 1. The two bodies are maintained at temperatures $T_1$ and $T_2$. To investigate the presence of multiple MPP and eliminate the influence of other materials on the results, the system is assumed to be situated in a vacuum environment with no substrate material present included in the system. The distances between the bodies and the graphene sheets adjacent to each other are represented as *d* and $d_{in}$, respectively. *d* and $d_{in}$ are selected as 10 nm for both, unless otherwise specified. A static quantized magnetic field *B* is applied perpendicularly to the system.

The radiative heat transfer coefficient (RHTC) is defined as follows to evaluate the NFRHT in the proposed system:

$$h = \frac{1}{4\pi^2} \int_0^\infty \hbar\omega \frac{\partial n}{\partial T} d\omega \int_0^\infty \xi(\omega,\kappa) \kappa d\kappa \tag{5}$$

where $\hbar$ is Planck's constant divided by $2\pi$, and $n = [\exp(\hbar\omega/k_B T)-1]^{-1}$ denotes the mean photon occupation number. $\xi(\omega,\kappa)$ is the energy transmission coefficient

$$\xi(\omega,\kappa) = \begin{cases} \mathrm{Tr}\left[\left(\mathbf{I}-\mathbf{R}_2^\dagger\mathbf{R}_2-\mathbf{T}_2^\dagger\mathbf{T}_2\right)\mathbf{D}\left(\mathbf{I}-\mathbf{R}_1\mathbf{R}_1^\dagger-\mathbf{T}_1\mathbf{T}_1^\dagger\right)\mathbf{D}^\dagger\right], \kappa < \kappa_0 \\ \mathrm{Tr}\left[\left(\mathbf{R}_2^\dagger-\mathbf{R}_2\right)\mathbf{D}\left(\mathbf{R}_1-\mathbf{R}_1^\dagger\right)\mathbf{D}^\dagger\right]e^{-2|\kappa_z|d}, \kappa > \kappa_0 \end{cases} \tag{6}$$

where $\kappa$, $\kappa_0 = \omega/c$, and $\kappa_z = \sqrt{\kappa_0^2 - \kappa^2}$ are the surface-parallel wave vector, the wave vector and the tangential wave vector in vacuum, respectively. $\mathbf{D} = \left(\mathbf{I}-\mathbf{R}_1\mathbf{R}_2 e^{2i\kappa_z d}\right)^{-1}$, and $\mathbf{R}_i$ is the reflection coefficient matrices, which are introduced in details in the following part.

The utilization of a generalized 4 × 4 T-matrix formalism for arbitrary anisotropic 2D layers is motivated by the anisotropic nature of graphene in the presence of a magnetic field. This method provides a detailed description of the physical characteristics of graphene in the presence of a magnetic field. This method differs from existing literature that utilizes the dielectric function of magneto-optical graphene, which may lead to physically unrealistic results due to simplifications [57]. General relations for the surface wave dispersions and the reflection matrix **R** are derived. Within the homogenization approach, the electromagnetic (EM) field response can be described by a fully populated conductivity tensor in the wave-vector space, denoted as [61],

$$\hat{\sigma}'' = \begin{pmatrix} \sigma''_{xx} & \sigma''_{xy} \\ \sigma''_{yx} & \sigma''_{yy} \end{pmatrix} = \frac{1}{k^2} \begin{pmatrix} k_x^2 \sigma_{xx} + k_y^2 \sigma_{yy} + k_x k_y \left( \sigma_{xy} + \sigma_{yx} \right) & k_x^2 \sigma_{xy} - k_y^2 \sigma_{yx} + k_x k_y \left( \sigma_{yy} - \sigma_{xx} \right) \\ k_x^2 \sigma_{yx} - k_y^2 \sigma_{xy} + k_x k_y \left( \sigma_{yy} - \sigma_{xx} \right) & k_x^2 \sigma_{yy} + k_y^2 \sigma_{xx} - k_x k_y \left( \sigma_{xy} + \sigma_{yx} \right) \end{pmatrix} \quad (7)$$

The 4 × 4 T-matrix is obtained by substituting p-wave and s-wave into the boundary conditions of the metasurface. This matrix establishes the relationship between the electric field components and magnetic field components in the media above and below the metasurface [62]:

$$\begin{pmatrix} E_{p1}^+ \\ E_{p1}^- \\ E_{s1}^+ \\ E_{s1}^- \end{pmatrix} = \hat{T}_{1 \to 2} \begin{pmatrix} E_{p2}^+ \\ E_{p2}^- \\ E_{s2}^+ \\ E_{s2}^- \end{pmatrix} \quad (8)$$

here, subscripts $p$ and $s$ represent the p-polarized (TM) and the s-polarized (TE), respectively, and the signs of + and - represent forward and backward waves, respectively. $\hat{T}_{1 \to 2}$ can be derived as:

$$\hat{T}_{1 \to 2} = \frac{1}{2} \begin{bmatrix} \frac{k_2 n_1}{\varepsilon_1 n_2} \begin{pmatrix} P_{12}^{++} & P_{12}^{-+} \\ P_{12}^{--} & P_{12}^{+-} \end{pmatrix} & \frac{n_1}{\varepsilon_1} \sigma_{xy} \begin{pmatrix} 1 & 1 \\ -1 & -1 \end{pmatrix} \\ \frac{k_2 \mu_1}{k_1 n_2} \sigma_{yx} \begin{pmatrix} 1 & 1 \\ -1 & -1 \end{pmatrix} & \frac{\mu_1}{k_1} \begin{pmatrix} S_{12}^{++} & S_{12}^{-+} \\ S_{12}^{--} & S_{12}^{+-} \end{pmatrix} \end{bmatrix} \quad (9)$$

where the p-waves components are [62]

$$P_{12}^{++} = \frac{\varepsilon_1}{k_1} + \frac{\varepsilon_2}{k_2} + \sigma_{xx}, \quad P_{12}^{-+} = \frac{\varepsilon_1}{k_1} - \frac{\varepsilon_2}{k_2} + \sigma_{xx}$$
$$P_{12}^{--} = \frac{\varepsilon_1}{k_1} - \frac{\varepsilon_2}{k_2} - \sigma_{xx}, \quad P_{12}^{+-} = \frac{\varepsilon_1}{k_1} + \frac{\varepsilon_2}{k_2} - \sigma_{xx} \quad (10)$$

and the s-waves components are [62]

$$S_{12}^{++} = \frac{k_1}{\mu_1} + \frac{k_2}{\mu_2} + \sigma_{yy}, \quad S_{12}^{-+} = \frac{k_1}{\mu_1} - \frac{k_2}{\mu_2} + \sigma_{yy}$$
$$S_{12}^{--} = \frac{k_1}{\mu_1} - \frac{k_2}{\mu_2} - \sigma_{yy}, \quad S_{12}^{+-} = \frac{k_1}{\mu_1} + \frac{k_2}{\mu_2} - \sigma_{yy} \quad (11)$$

here, subscripts 1 and 2 represent the media above and below the metasurface, respectively. $k_z$, $\mu$, and $\varepsilon$ are tangential wavevector along $z$ direction in media, magnetic permeability of media, and dielectric function of media. $\varepsilon_0$ is the vacuum permittivity, and $\mu_0$ is the vacuum permeability. In general, for any 4 × 4 T-matrix that links all the electric field components in a first layer with those in $N$ layer [62],

$$\begin{pmatrix} E_{p1}^+ \\ E_{p1}^- \\ E_{s1}^+ \\ E_{s1}^- \end{pmatrix} = \begin{pmatrix} T_{11} & T_{12} & T_{13} & T_{14} \\ T_{21} & T_{22} & T_{23} & T_{24} \\ T_{31} & T_{32} & T_{33} & T_{34} \\ T_{41} & T_{42} & T_{43} & T_{44} \end{pmatrix} \begin{pmatrix} E_{pN}^+ \\ E_{pN}^- \\ E_{sN}^+ \\ E_{sN}^- \end{pmatrix} \quad (12)$$

The reflection matrix **R** defined and expressed in terms of the T-matrix elements are as follows [62]:

$$\begin{aligned} r_{pp} &= \left.\frac{E_{p1}^-}{E_{p1}^+}\right|_{E_{s1}^+=0} = \frac{T_{21}T_{33} - T_{23}T_{31}}{T_{11}T_{33} - T_{13}T_{31}} \\ r_{ps} &= \left.\frac{E_{s1}^-}{E_{p1}^+}\right|_{E_{s1}^+=0} = \frac{T_{41}T_{33} - T_{43}T_{31}}{T_{11}T_{33} - T_{13}T_{31}} \\ r_{sp} &= \left.\frac{E_{p1}^-}{E_{s1}^+}\right|_{E_{p1}^+=0} = \frac{T_{11}T_{23} - T_{13}T_{21}}{T_{11}T_{33} - T_{13}T_{31}} \\ r_{ss} &= \left.\frac{E_{s1}^-}{E_{s1}^+}\right|_{E_{p1}^+=0} = \frac{T_{11}T_{43} - T_{13}T_{41}}{T_{11}T_{33} - T_{13}T_{31}} \end{aligned} \quad (13)$$

The transmission matrix **T** defined and expressed in terms of the T-matrix elements are as follows [62]:

$$\begin{aligned} t_{pp} &= \left.\frac{E_{pN}^+}{E_{p1}^+}\right|_{E_{s1}^+=0} = \frac{T_{33}}{T_{11}T_{33} - T_{13}T_{31}} \\ t_{ps} &= \left.\frac{E_{sN}^+}{E_{p1}^+}\right|_{E_{s1}^+=0} = \frac{-T_{31}}{T_{11}T_{33} - T_{13}T_{31}} \\ t_{sp} &= \left.\frac{E_{pN}^+}{E_{s1}^+}\right|_{E_{p1}^+=0} = \frac{-T_{13}}{T_{11}T_{33} - T_{13}T_{31}} \\ t_{ss} &= \left.\frac{E_{sN}^+}{E_{s1}^+}\right|_{E_{p1}^+=0} = \frac{T_{11}}{T_{11}T_{33} - T_{13}T_{31}} \end{aligned} \quad (14)$$

The formalism developed above can be easily generalized for arbitrary multilayer system by multiplying the T-matrices for each layer.

## III. RESULTS AND DISCUSSION

### A. Modulation and Physical Mechanism of Multiple MPP in NFRHT

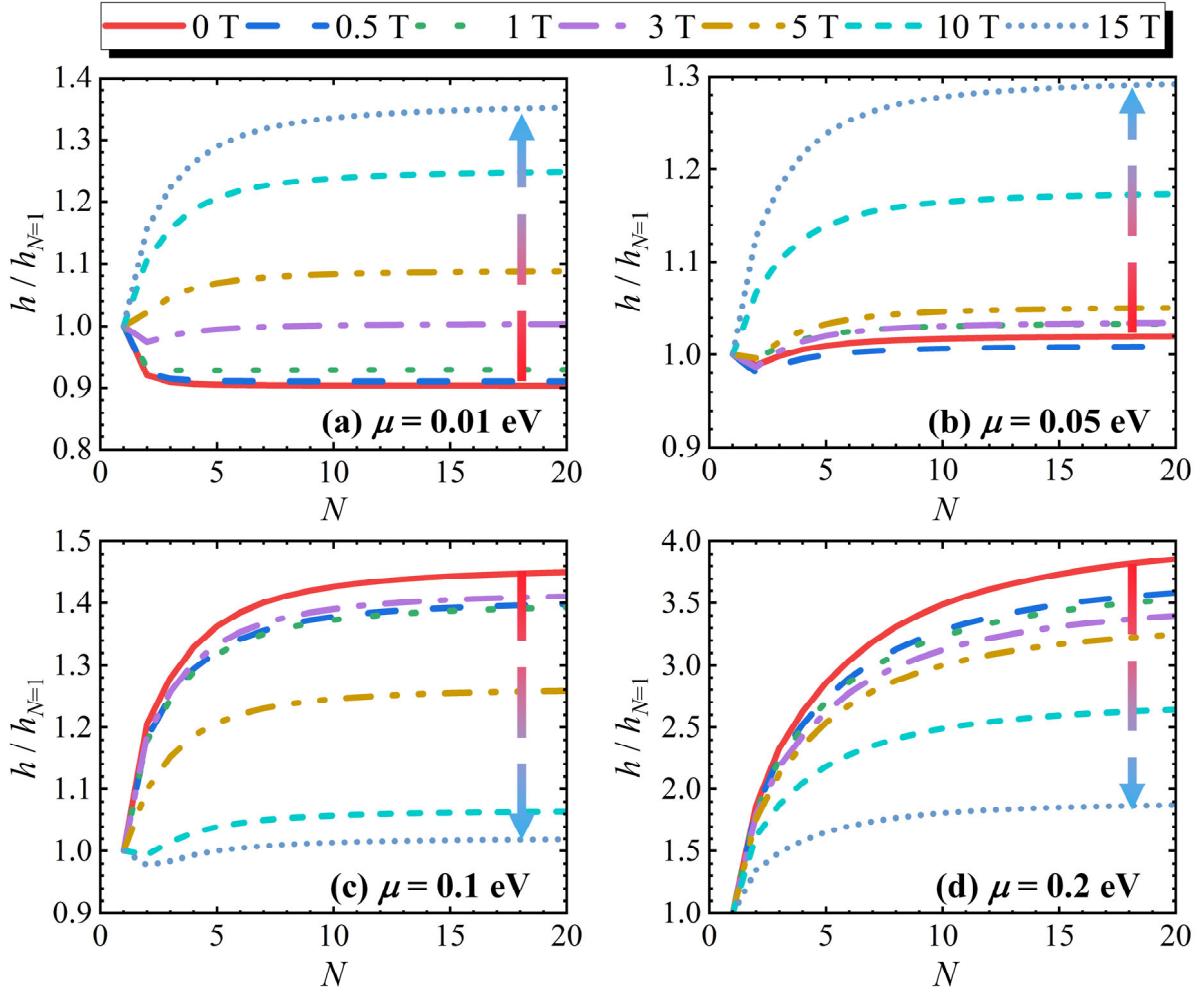

Fig. 2. The ratio of RHTC between multilayered graphene sheets to one layer system for different chemical potentials of graphene : (a) 0.01 eV, (b) 0.05 eV, (c) 0.1 eV, and (d) 0.2 eV at different magnitudes of magnetic fields and number of layers.

This section focuses on the modulation of NFRHT by multiple MPP and explores their physical origin and evolution. Previous studies have reported strong mediation of NFRHT between closely separated graphene sheets through thermally excited SPP modes [41]. The key parameter under consideration is the number of graphene sheets, and we compare the results for $N > 1$ with those for $N = 1$ to demonstrate the distinction between single MPP and multiple MPP scenarios. We present plots of the ratio of RHTC between multilayered graphene sheets and a single-layer system for graphene with different chemical potentials $\mu = 0.01$ eV, 0.05 eV, 0.1 eV, 0.2 eV in Figs. 2(a)-(d), respectively. The chemical potentials of graphene examined in this study are experimentally achievable [63]. The red curves represent the results without a magnetic field, and it is important to note that multiple SPPs exist in NFRHT only in the absence of a magnetic field [47]. The arrows in the figure indicate the transition from the zero-field curve to the curve corresponding to $B = 15$ T magnetic field. For $\mu = 0.01$ eV, the ratio decreases with increasing $N$ for $B < 3$ T (including zero field), but it increases with $N$ for $B > 5$ T. This indicates a distinct modulation effect in NFRHT between multiple SPPs and multiple MPP at this chemical potential. For $\mu = 0.05$ eV, the ratios consistently above unity for different numbers of graphene sheets, indicating that multiple MPP enhance NFRTH at this chemical potential. However, the results are significantly different for $\mu = 0.1$ eV and $\mu = 0.2$ eV in Figs. 2(c)-(d), where the two arrows point in the opposite direction compared to Figs. 2(a)-(b). The ratios gradually decrease with increasing magnetic field intensity. It is important to note that even under such chemical potentials, the presence of multiple MPP can still enhance NFRHT, although this enhancement is weaker compared to that mediated by multiple SPPs. These findings lead us to the conclusion that multiple MPP can either decrease or increase NFRHT, depending on the strength of the magnetic field and the chemical potentials of graphene. Since the system does not involve any additional materials, the modulation effect of multiple MPP on NFRHT is more pronounced compared to graphene-based hyperbolic metamaterials [57]. The results indicate that, in comparison to multiple SPPs, the multilayered modulation effect on NFRHT driven by multiple MPP exhibits enhancement at low chemical potentials of graphene and a decline at relatively high chemical potentials.

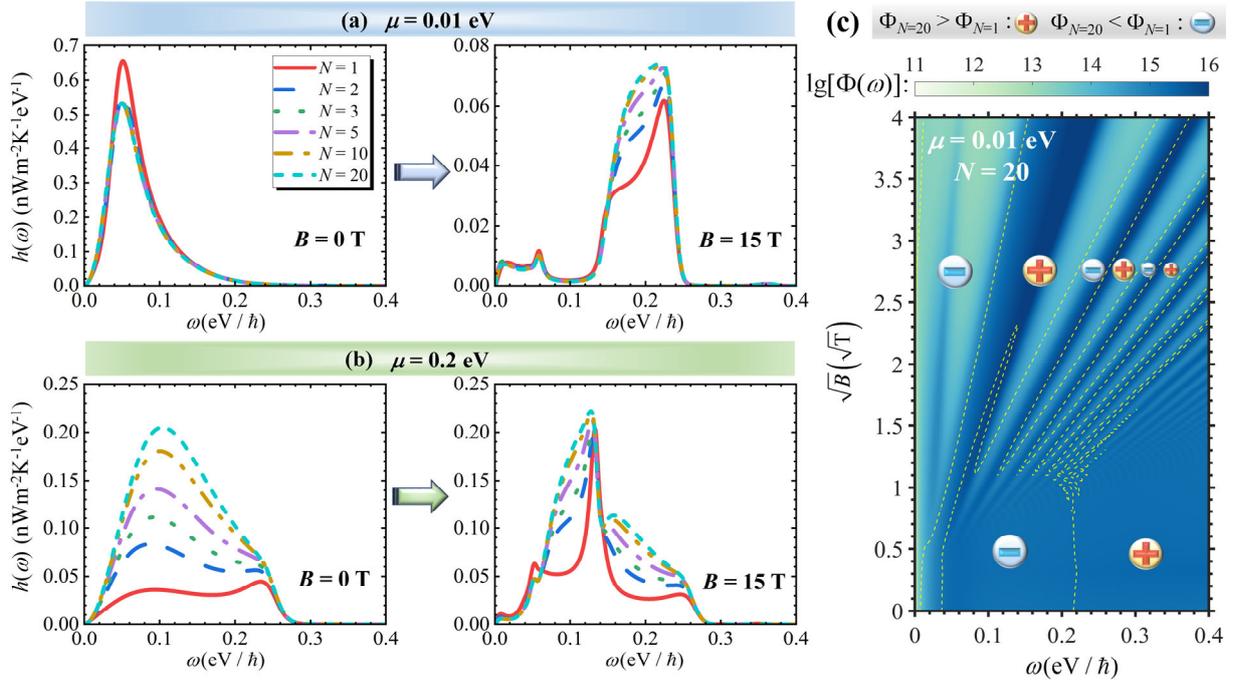

Fig. 3. The spectral RHTC for different numbers of graphene sheets with the chemical potentials of graphene (a) $\mu = 0.01$ eV and (b) $\mu = 0.2$ eV. (c) The spectral tranfer function for $N = 20$ multilayered graphene sheets with $\mu = 0.01$ eV at different magnitudes of magnetic fields, the enhancement and attenuation compared to $N = 1$ is denoted with the plus and minus signs, respectively.

Why does the modulation effect of multiple MPP in NFRHT exhibit a fundamentally different trend compared to multiple SPPs with varying magnitudes of graphene's chemical potentials? To investigate the underlying mechanism behind the effects driven by multiple MPP, we plot the spectral RHTC for graphene with chemical potentials $\mu = 0.01$ eV and $\mu = 0.2$ eV in Figs. 3(a) and (b), respectively, considering various numbers of graphene sheets. The figures depict the spectral RHTC for both zero magnetic field and a magnetic field of $B = 15$ T. At $\mu = 0.01$ eV, the spectral RHTC shows attenuation caused by multiple SPPs under zero magnetic field, whereas an enhancement of the spectral RHTC occurs due to multiple MPP when a magnetic field is present. For the higher chemical potential of $\mu = 0.2$ eV, both the curves for zero field and $B = 15$ T exhibit an enhancement in the spectral RHTC when multiple SPPs and MPP are employed. It is clearly observed that the enhancement effect resulting from multiple MPP is slightly weaker compared to that driven by multiple SPPs.

The modulation effects of multiple SPPs and multiple MPP in NFRHT with $\mu = 0.01$ eV exhibit fundamentally different behaviors, as depicted in Fig. 2(a) and further illustrated in Fig. 3(a).

In order to gain a comprehensive understanding of the evolution of multiple MPP, we plot the transfer function Φ(ω) as a function of magnetic field and frequency. The transfer function is defined as the integral of energy transmission coefficients over wave vectors, as described in Eq. (5). Fig. 3(c) displays the corresponding results for $\mu$ = 0.01 eV and $N$ = 20. The enhancements and attenuations of the transfer function, relative to $N$ = 1, are indicated by plus and minus signs, respectively. It is evident that, under weak magnetic fields ($B$ < 1 T), the intraband transition at low frequencies is weakened, while the first interband transition at high frequencies is enhanced. The attenuation at low frequencies plays a vital role in radiative heat transfer, leading to a reduction in NFRHT compared to a single-layer system. With an increase in $B$, the transfer function displays oscillatory behavior resembling Shubnikov–de Haas oscillations in the spectrum [54]. The intraband transition nearly vanishes, and the first interband transition becomes the dominant contribution. The peak of the transfer function primarily resides near the first interband transition, and it surpasses that of the single-layer system within this frequency range. Consequently, the NFRHT is enhanced by multiple MPP relative to a single-layer configuration under these circumstances [53, 54].

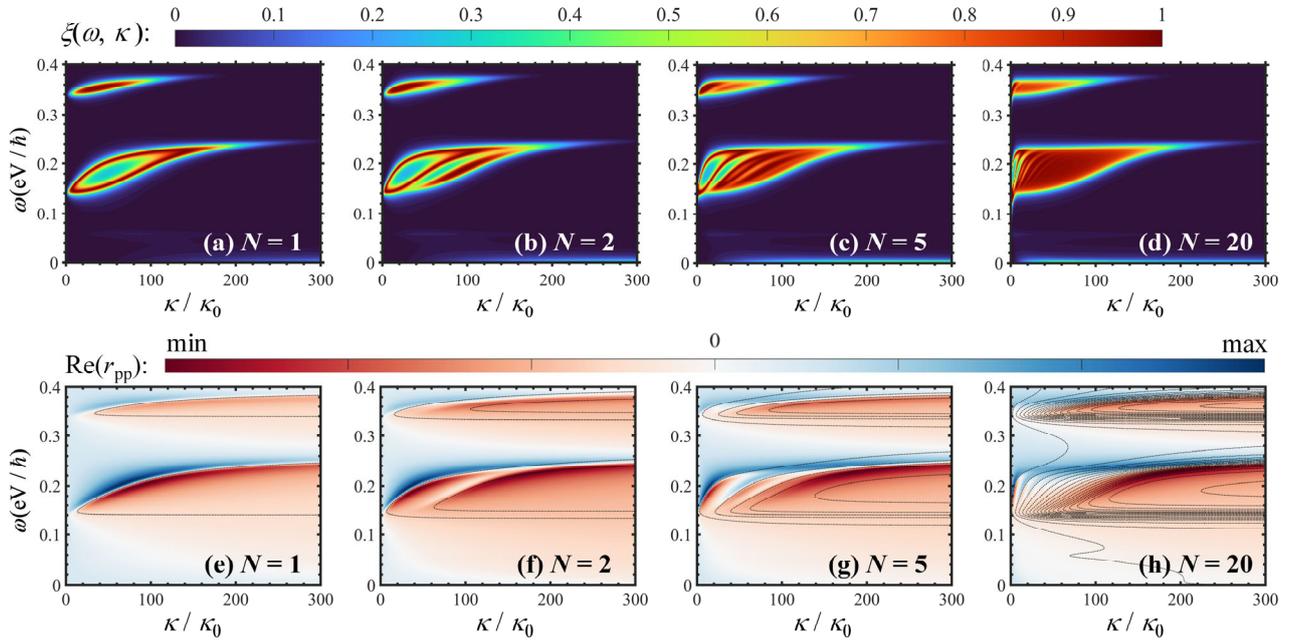

Fig. 4. With $\mu$ = 0.01 eV and $B$ = 15 T, the energy transmission coefficients $\xi(\omega, \kappa)$ for different numbers of graphene sheets: (a) $N$ = 1, (b) $N$ = 2, (c) $N$ = 5, (d) $N$ = 20; and the real parts of reflection coefficients Re ($r_{pp}$): (e) $N$ = 1, (f) $N$ = 2, (g) $N$ = 5, (h) $N$ = 20. The dispersion relations are illustrated with the dashed lines.

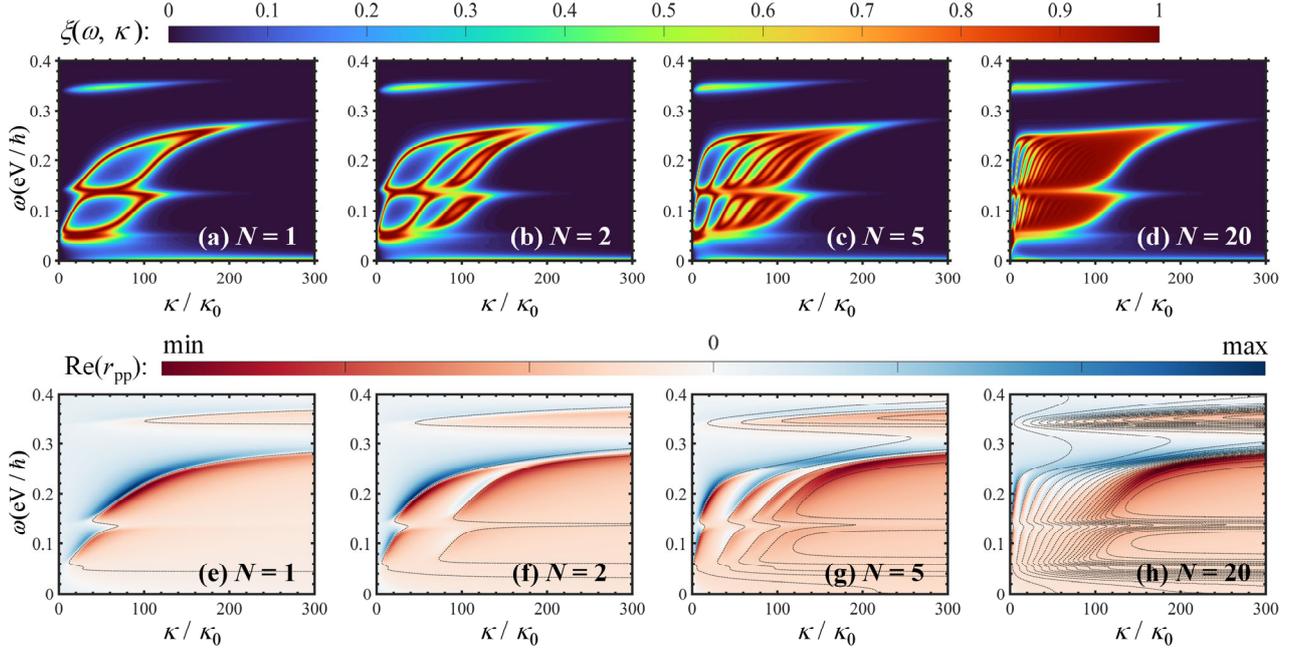

Fig. 5. With $\mu = 0.2$ eV and $B = 15$ T, the energy transmission coefficients $\xi(\omega, \kappa)$ for different numbers of graphene sheets: (a) $N = 1$, (b) $N = 2$, (c) $N = 5$, (d) $N = 20$; and the real parts of reflection coefficients Re ($r_{pp}$): (e) $N = 1$, (f) $N = 2$, (g) $N = 5$, (h) $N = 20$. The dispersion relations are illustrated with the dashed lines.

The above results suggest that the modulation of NFRHT can be achieved by varying the number of graphene sheets, and this modulation is linked to the evolution of multiple MPP. To elucidate the physical mechanism and evolution of multiple MPP in relation to the number of graphene sheets, we analyze the evolution of energy transmission coefficients and reflection coefficients in the considered multilayer structure. The energy transmission coefficients $\xi(\omega, \kappa)$ for different number of graphene sheets and the real parts of reflection coefficients Re ($r_{pp}$) are both given in Figs. 4 and 5 for $\mu = 0.01$ eV and $\mu = 0.2$ eV in the presence of field $B = 15$ T, respectively. The coupling between multiple MPP modes is most pronounced when both multilayered graphene structures share identical parameters, resulting in identical individual dispersions [41]. The dispersion of the combined system shows different numbers of dominant branches in the near-field spectral transfer. The dispersion relations of the surface characteristics are depicted by the black-dashed lines together with the reflection coefficients. For $\mu = 0.01$ eV, the two branches of MPP modes of the single graphene sheet denote the first and second interband transitions, respectively. As the number of graphene sheets increases, the original single curve of dispersion relation transforms to two, five and twenty curves [Figs. 4(e)-(h)]. Moreover, the number of MPP branches is also increasing, as shown in Figs. 4(a)-(d), and the different branches couple with each other intensely. The newly emerging multiple MPP branches occupy the region surrounded by the original MPP branches with $N = 1$ and expand into a larger area in the ($\omega$, $\kappa$) space. Considering the higher chemical potential of graphene ($\mu = 0.2$ eV), the two branches of MPP in Fig. 5(a) represent the intraband and first interband transitions, respectively. As $N$ increases, the MPP convert to multiple MPP, and the newly emerging branches of multiple MPP occupy the hollow area in the ($\omega$, $\kappa$) space. The above results indicate that, regardless of whether $\mu = 0.01$ eV or $\mu = 0.2$ eV, the enhancement of multiple MPP compared to single MPP is mainly attributed to the filling effect in the wave vector dimension. The multiple MPP are hardly influenced in the frequency dimension, which is consisent with the transfer function results in Figs. 3(a) and (b). The evolution from a single MPP to multiple MPP is entirely different from the evolution of SPPs, which is a simple progression from two to multiple branches with continuous contours in the ($\omega$, $\kappa$) space [47].

## B. Thermal Magnetoresistance Driven by Multiple MPP

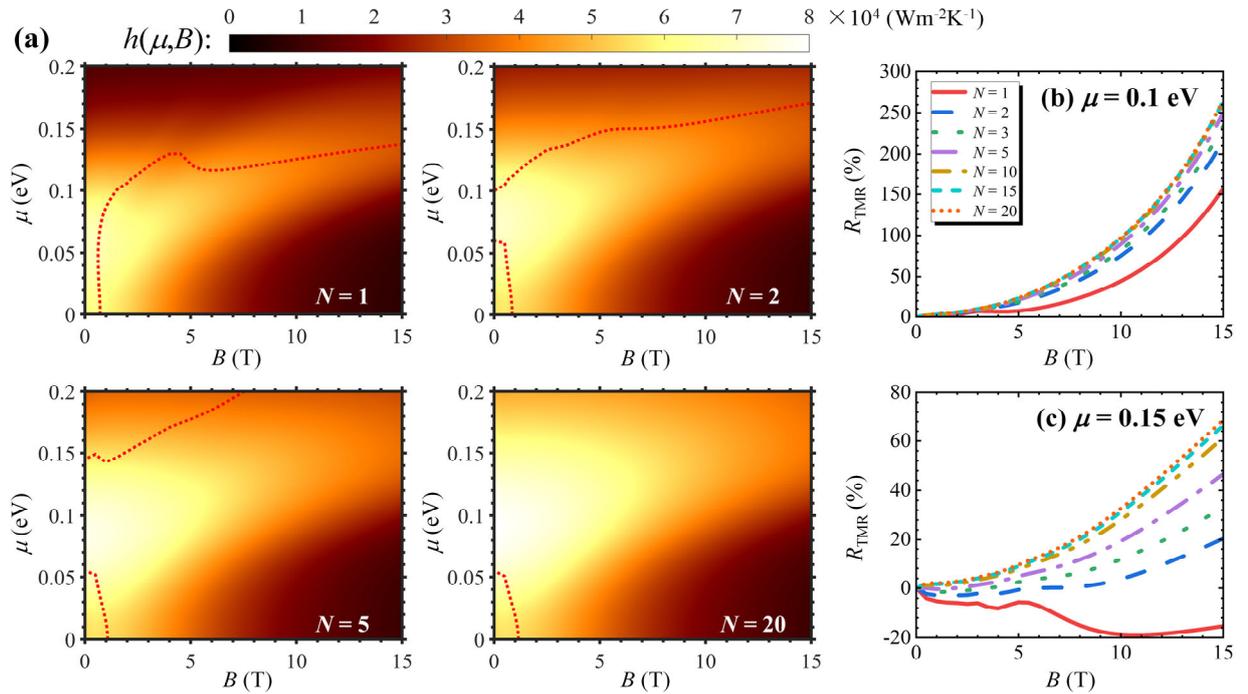

Fig. 6. (a) For different numbers of graphene sheets, the RHTC corresponding to different magnitudes of magnetic fields and chemical potentials of graphene. The relative thermal magnetoresistance ratio $R_{\text{TMR}} = [R(B)-R(0)]/R(0)=[h(0)/h(B)-1]\times 100\%$ for (b) $\mu = 0.1$ eV and (c) $\mu = 0.15$ eV.

It is evident from Eqs. (2)-(4) that the magneto-optical conductivity of graphene is influenced by both the intensity of magnetic fields and the chemical potentials of graphene. Under a specific chemical potential, the evolution of intraband and interband transitions is primarily governed by the intensity of magnetic fields [52]. The magnetic field intensity can modulate the Dirac energy spectrum of magneto-optical graphene, thus fundamentally impacting the characteristics of MPP. In this study, we examine the influence of magnetic field intensity on multiple MPP in the modulation of NFRHT. Fig. 6(a) presents the RHTC for various combinations of magnetic fields and chemical potentials of graphene, considering different numbers of graphene sheets. The red-dashed line represents the RHTC identical to that in the absence of magnetic fields. The results indicate that with an increasing number of graphene sheets, the trend of RHTC variation with respect to ($B$, $\mu$) gradually evolves. The evolution of the red-dashed line demonstrates that there is nearly no region of heat transfer attenuation compared to the zero-field case for large $N$. This suggests that multiple MPP primarily enhance NFRHT compared to multiple SPPs in most combinations of ($B$, $\mu$).

The discovery of the giant magnetoresistance effect by Grünberg and Fert in 1988 is regarded as a highly significant advancement in solid-state physics [64]. Building on this unique effect, a thermal analog called the giant thermal magnetoresistance effect is predicted in magneto-optical plasmonic structures within the domain of thermal photon heat transfer [36]. The relative thermal magnetoresistance ratio, denoted as $R_{TMR}$, is defined as $R_{TMR}=[R(B)-R(0)]/R(0)=[h(0)/h(B)-1]\times 100\%$. Here, $R$ represents the thermal magnetoresistance and is given by $R=1/h$. According to the definition, positive and negative values of $R_{TMR}$ indicate reduced and enhanced heat transfer, respectively, in comparison to the zero-field case. Figs. 6(b) and (c) illustrate the relative thermal magnetoresistance ratio for $\mu = 0.1$ eV and $\mu = 0.15$ eV, respectively. The lines represent the results for varying numbers of graphene sheets, with the red line corresponding to the single-layer system. When the chemical potential is $\mu = 0.1$ eV, $R_{TMR}$ is consistently positive and exhibits a monotonically increasing trend with the magnetic field magnitude. Furthermore, the values of $R_{TMR}$ become larger as the number of graphene sheets increases. The $R_{TMR}$ increases from 160 % in the case of a single-layer system to 265 % for a multilayered system with $N = 15$ (or 20) layers, under a magnetic field of $B = 15$ T. However, the results exhibit different trends for $\mu = 0.15$ eV in Fig. 6(c). In the case of a single-layer system, the values of RTMR are consistently negative for varying magnitudes of magnetic fields. However, for the multilayered system, the values of $R_{TMR}$ are consistently positive and increase with the number of layers ($N$). Specifically, $R_{TMR}$ increases from -16% for the single-layer system to 68% for the $N = 20$ multilayered system under a magnetic field of $B = 15$ T. The above results indicate that the multiple MPP are significantly influenced by the intensity of magnetic fields in the modulation of NFRHT.

Furthermore, the relative thermal magnetoresistance ratio observed in this study is higher than those obtained in graphene-based hyperbolic metamaterials, which consist of multilayered structures alternating between graphene and polar dielectric materials [57]. Therefore, we can conclude that achieving a higher thermal magnetoresistance driven by multiple MPP does not require the inclusion of additional materials in the system. This observation is attributed to the pronounced spectral dependence of localized multiple MPP on the magnetic field magnitude. Furthermore, when compared to existing literature on thermal magnetoresistance [36, 37], the results obtained in this study using multiple MPP in graphene demonstrate exceptional performance despite the relatively simple structure employed [65].

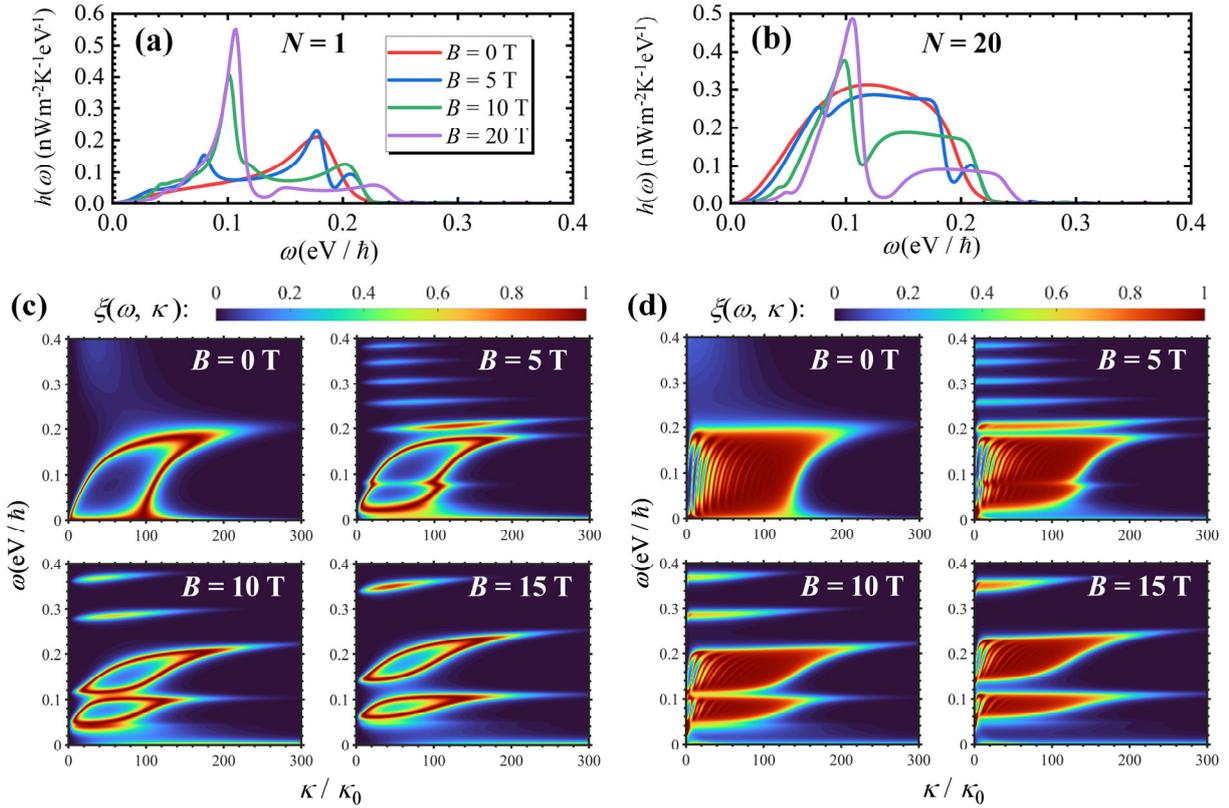

Fig. 7. With $\mu = 0.15$ eV, the spectral RHTC for (a) $N = 1$ and (b) $N = 20$ at different magnitudes of magnetic field. The energy transmission coefficients for (c) $N = 1$ and (d) $N = 20$.

The observed transition from negative to positive values of the relative thermal magnetoresistance ratio in Fig. 6(c) represents a novel and intriguing physical phenomenon, with potential implications for enhancing the flexibility of manipulating the NFRHT. To gain a comprehensive understanding of this unique phenomenon and the evolution of multiple MPP in response to varying magnetic field intensities, we present in Figs. 7(a) and (b) the spectral RHTC for $N = 1$ and $N = 20$, respectively, with $\mu = 0.15$ eV. Additionally, Figs. 7(c) and (d) depict the energy transmission coefficients at various magnetic field magnitudes $N = 1$ and $N = 20$, respectively. The results presented in Figs. 7(a) and (c) illustrate the evolution of MPP, whereas those shown in Figs. 7(b) and (d) depict the evolution of multiple MPP in relation to magnetic field intensity. In the single-layer system, as the magnetic field increases from zero to $B = 5$ T, 10 T, and 15 T, the low-frequency symmetric and high-frequency antisymmetric branches of SPPs progressively transform into distinct branches of MPP due to interband and intraband transitions among different Landau levels. Due to the Dirac-like nature of quasiparticle excitations in graphene, the magneto-optical conductivities of graphene demonstrate highly atypical behavior in relation to frequency, chemical potential, and applied field [59]. Consequently, the unimodal curve transforms into multi-peak curves exhibiting Shubnikov–de Haas-like oscillations in the spectral heat flux. This suggests that the negative RTMR observed in Fig. 6(c) can be attributed to the heat flux enhancement resulting from intraband transitions at low frequency in the presence of magnetic fields.

In the case of the $N = 20$ multilayered system, the contours of energy transmission coefficients encompass the entire ($\omega$, $\kappa$) space in the absence of a magnetic field, resulting from the influence of multiple SPPs. Upon the application of a magnetic field in the system, the multiple SPPs undergo a transformation into multiple MPP. The interconnected branches of these MPP continue to converge as the magnetic field intensity increases. However, the initial entire area occupied by multiple SPPs gradually evolves into distinct regions. The segregation of multiple MPP results in a narrower frequency range occupied by the energy transmission coefficients, thereby weakening the NFRHT. Consequently, the positive $R_{TMR}$ observed in Figs. 6(b) and (c) can be ascribed to the distinct effect of multiple MPP.

## C. Effect of Structure Parameters on Multiple MPP

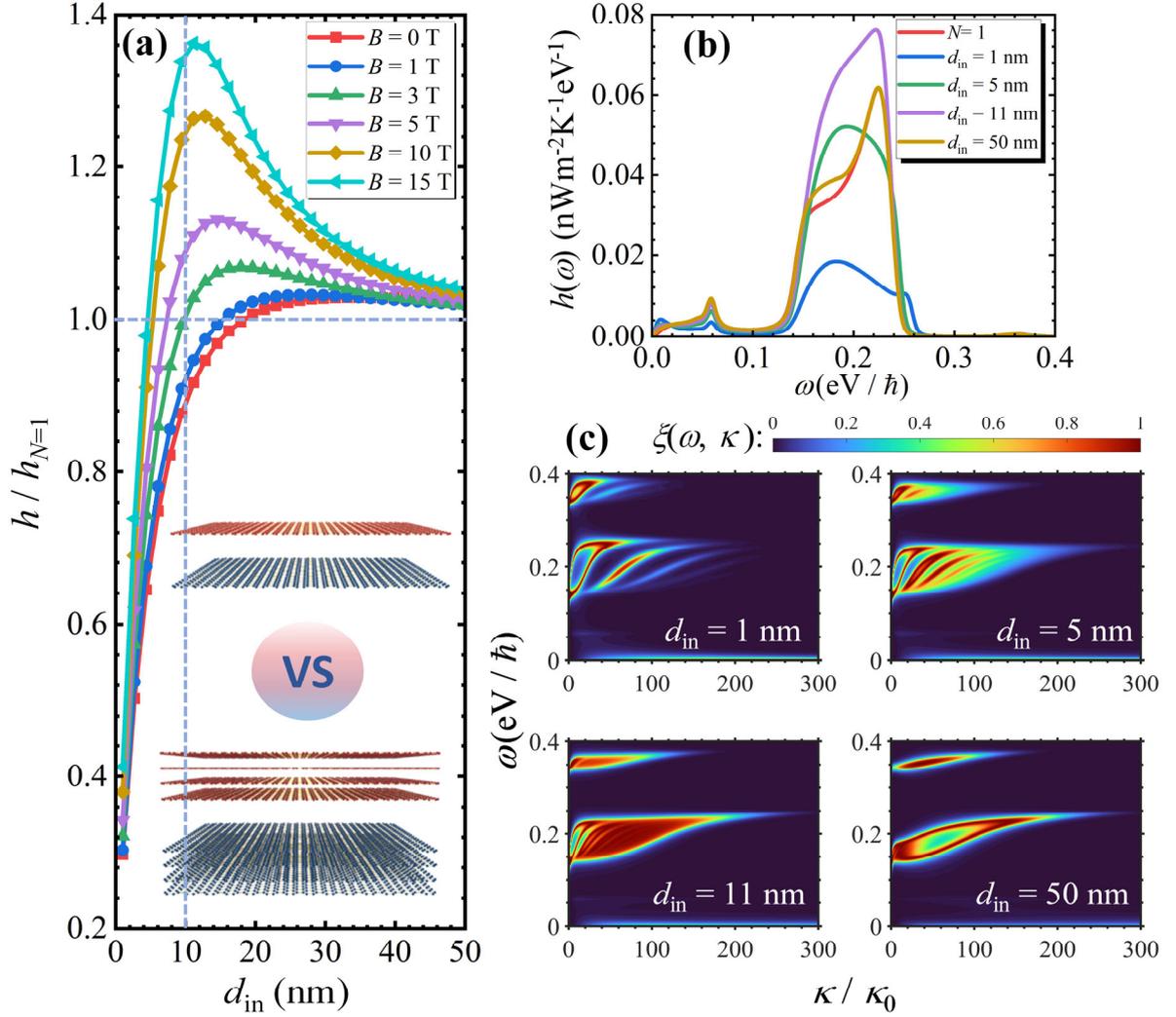

Fig. 8  With $\mu = 0.01$ eV, (a) the ration of RHTC between $N = 10$ multilayered system to that of $N = 1$ at different separation distance between adjacent graphene sheets with a constant value of $d = 10$ nm. (b) The spectral RHTC and (c) energy transmission coefficients corresponding to different values of $d_{in}$.

The key structural parameters in the current configuration are the separation distances between the two bodies and the adjacent graphene sheets, labeled as $d$ and $d_{in}$, respectively. The separation distance between adjacent graphene sheets can be regarded as the density of the graphene system, influencing the occurrence of multiple MPP through the modulation of surface characteristics and reflection coefficients of the two bodies. In order to illustrate the impact of the body's compactness on multiple MPP in the modulation of NFRHT, we present the plot in Fig. 8(a) showing the ratio of RHTC between an $N = 10$ multilayered system and an $N = 1$ system at varying separation distances between adjacent graphene sheets, with a fixed value of d = 10 nm. Each curve represents the results obtained for different magnitudes of magnetic fields. The figure includes two dashed lines representing $h/h_{N=1} = 1$ and $d_{in} = 10$ nm. In the absence of a magnetic field in the system, the results indicate a weakened heat transfer due to multiple SPPs with smaller din compared to the $N = 1$ system. Additionally, the RHTC increases with $d_{in}$ until reaching a constant value at larger $d_{in}$. The curve corresponding to the absence of a magnetic field shows a monotonic trend. Upon the application of a magnetic field to the system, the curves exhibit a dramatic increase with $d_{in}$ at smaller values, especially for strong magnetic fields. These curves demonstrate non-monotonic trends with $d_{in}$, reaching a peak near the position of $d = d_{in}$. This suggests that multiple MPP can significantly enhance the NFRHT by adjusting the density of the multilayered graphene sheets. Moreover, as the magnetic field strength increases, the peak values also increase. Nevertheless, all the curves eventually decrease to the same value at larger $d_{in}$. This decrease is mainly due to the decoupling of multiple MPP and SPPs when the graphene sheets are sufficiently separated.

To investigate the underlying mechanism responsible for the observed peaks in Fig. 8(a), we plotted the spectral RHTC and energy transmission coefficients for various values of $d_{in}$ in Figs. 8(b) and (c), respectively. The trend of heat flux variation with respect to $d_{in}$ is clearly demonstrated in the spectral space. Specifically, the line corresponding to $d_{in} = 11$ nm exhibits significant enhancement compared to the other parameters. Fig. 8(c) illustrates that the energy transmission coefficients reveal the separation of different branches of multiple MPP at $d_{in} = 1$ nm. As $d_{in}$ increases to 5 nm, the branches of multiple MPP gradually converge, and the modes become stronger. When $d_{in}$ increases to 11 nm, the branches of multiple MPP converge and nearly form a unified structure. When the separation distance between adjacent graphene sheets closely matches that between the two bodies, the interactions between the sheets display a pronounced resonance effect. The intense resonance effect primarily arises from the alignment of geometric parameters, enabling the coupling between adjacent graphene sheets to align with the coupling relations between the two bodies. As $d_{in}$ increases to 50 nm, both the resonance effect and the multiple branches diminish, primarily due to the decoupling of multiple MPP caused by the attenuation of modes over long distances. When $d_{in}$ becomes excessively large, only a few graphene sheets in proximity to the interface between the two bodies contribute to heat transfer, while the sheets far away from the interface have limited interaction. These findings and principles establish a framework for enhancing NFRHT with multiple

MPP.

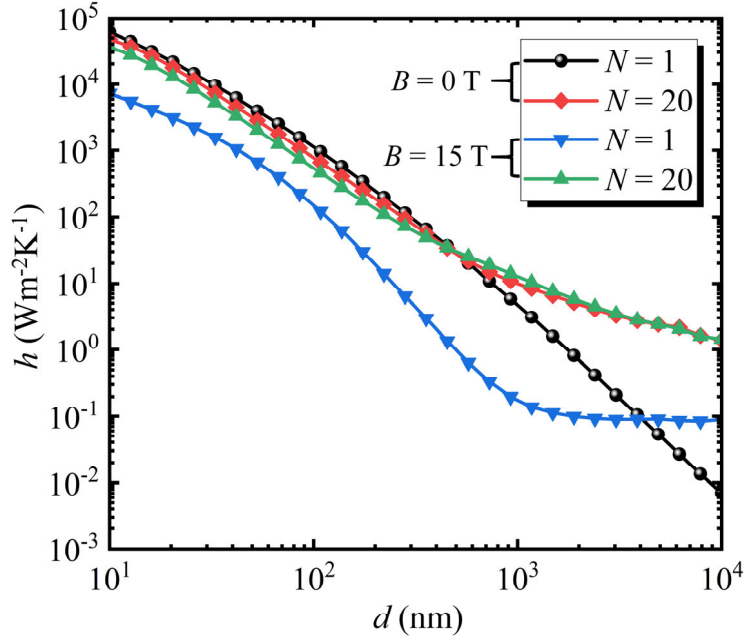

Fig. 9. With $\mu = 0.01$ eV, the RHTC for $N = 1$ and $N = 20$ system at different separation distance $d$ (with constant $d_{in} = 10$ nm) between the two bodies in the absence and presence of magnetic fields ($B = 15$ T).

Another structure parameter involved in the present structure is the separation distance between the two bodies, denoted as $d$ in Fig. 1. Here in Fig. 9, we plot the RHTC for $N = 1$ and $N = 20$ system at different $d$ (with constant $d_{in} = 10$ nm) in the absence and presence of magnetic fields ($B = 15$ T), respectively. In the absence of a magnetic field in the system, the differences between the results for the $N = 1$ and $N = 20$ systems are attributed to the presence of multiple SPPs. The results indicate that heat transfer is attenuated in the near-field regime, while it is significantly enhanced in the far-field regime. The results exhibit significant differences between the two curves of $N = 1$ and $N = 20$ in the presence of a magnetic field. The multiple MPP enhance the heat transfer this time, no matter in the near-field or the far-field regime. An interesting phenomenon is observed: the two curves for $N = 1$ exhibit two distinct trends, particularly in the far-field, which are influenced by the presence of SPPs and multiple SPPs. However, the curves for $N = 20$ in the absence and presence of magnetic fields are almost coincident in the far-field regime. It means that the multiple MPP and multiple SPPs are nearly the same at large separation distance. This is mainly attributed to the attenuation and decoupling of multiple modes due to the large distance.

## IV. Conclusions

In this work, we investigate the impact of multiple magnetoplasmon polaritons on the near-field radiative heat transfer between two structures composed of multilayered graphene sheets under an external magnetic field. We clarify the mechanism and evolution of multiple MPP stemming from multilayered magneto-optical graphene sheets. Multiple MPP are utilized to mediate, enhance, and tune the NFRHT through reasonable variations in the properties of graphene, the number of graphene sheets, the intensity of the magnetic field, and the geometric parameters of the system. Specifically, we have demonstrated significant differences between multiple magnetoplasmon polaritons and single MPP or multiple SPPs in modulating and manipulating NFRHT, particularly when the chemical potentials of graphene are small (0.01 eV). This behavior arises from the coupling between the significant contributions of surface states at multiple surfaces and Shubnikov–de Haas-like oscillations in the spectrum, indicating a transformation of intraband and interband transitions. In comparison to multiple SPPs, the multilayered modulation effect on NFRHT driven by multiple MPP, exhibits an enhancement at low chemical potentials of graphene, and a decline at relatively high chemical potentials.

The proposed system also predicts a giant thermal magnetoresistance effect, with the relative thermal magnetoresistance ratio increasing from 160% in a single-layer system to 265% in a multilayered system. Furthermore, we demonstrate the existence of a transition from negative to positive values of the relative thermal magnetoresistance ratio, ranging from -16% in a single-layer system to 68% in this multilayered system. Additionally, we have discovered that when the separation distance between adjacent graphene sheets is approximately the same as that between the two bodies, heat transfer reaches its maximum. We demonstrate that this remarkable behavior is attributed to the strong resonance effect resulting from the interactions between different graphene sheets at these geometric parameters. The findings presented in this study provide insights into thermal photon-based communication technology and hold the potential for a magnetically controllable thermal switch.

**Declaration of competing interest**

The authors declare that they have no known competing financial interests or personal relationships that could have appeared to influence the work reported in this paper.

**Data availability**

Data will be made available on request.

**Acknowledgements**

The supports of this work by the National Natural Science Foundation of China (No. 52206082, No. 52276055), China Postdoctoral Science Foundation (No. 2021TQ0086), the Natural Science Foundation of Heilongjiang Province (No. LH2022E063), Postdoctoral Science Foundation of Heilongjiang Province (No. LBH-Z21013), Excellent Thesis of Masters and Doctors of New Era Heilongjiang Province (No. LJYXL2022-009) are gratefully acknowledged.